\documentclass[prb,aps,twocolumn,superscriptaddress,showpacs,10pt,floatfix]{
revtex4-1}
\usepackage{amsmath}
\usepackage{amsfonts}
\usepackage{amssymb}
\usepackage{bm}
\usepackage{graphicx}
\usepackage{datetime}
\usepackage{color}
\begin{document}
%%%%%%%%%%%%%%%%%%%%%%%%%%%%%%%%%%%%%%%%%%%%%%%%%%%%%%%%%%%%
% comment out for single column
%\twocolumn[%
%\hsize\textwidth\columnwidth\hsize\csname@twocolumnfalse\endcsname
%\draft
%%%%%%%%%%%%%%%%%%%%%%%%%%%%%%%%%%%%%%%%%%%%%%%%%%%%%%%%%%%%
\newcommand{\figwidth}{0.95\columnwidth}
\newcommand{\ffigwidth}{0.4\columnwidth}
%%%affiliations
\newcommand{\warwick}{Department of Physics, University of Warwick, Coventry,
CV4 7AL, United Kingdom}
\newcommand{\warwickcsc}{Centre for Scientific Computing, University of Warwick,
Coventry, CV4 7AL, United Kingdom}
\newcommand{\unicamp}{Instituto de Fisica Gleb Wataghin, Universidade Estadual
de Campinas,
Rua S\'{e}rgio Buarque de Holanda, 777 Cidade Universit\'{a}ria 13083-859
Campinas, SP Brazil}
\newcommand{\unicamplimeira}{Faculdade de Ci\^{e}ncias Aplicadas, Universidade
Estadual de Campinas, 13484-350 Limeira, SP Brazil}

\title{Robust signatures in the current-voltage characteristics of DNA molecules oriented between two graphene nanoribbon electrodes}

\author{Carlos J. P\'aez}
\affiliation{\unicamp}
%\email[Corresponding author:]{cjpaezg@ifi.unicamp.br}
\author{Peter A. Schulz}
\affiliation{\unicamp}
\affiliation{\unicamplimeira}
\author{Neil Wilson}
\affiliation{\warwick}
\author{Rudolf A. R\"omer}
\affiliation{\warwick} 
\affiliation{\warwickcsc}

\date{1.69, compiled \today, \currenttime}
%\date{\today}
%
\begin{abstract}
In this work we numerically calculate the electric current through three kinds
of DNA sequences (telomeric, $\lambda$-DNA, and p53-DNA) described by different
heuristic models. A bias voltage is applied between two zig-zag edged graphene
contacts attached to the DNA segments, while a gate terminal modulates the
conductance of the molecule. The calculation of current is performed by
integrating
the  transmission function (calculated using the lattice Green's function) over
the range of energies allowed by the chemical potentials.  We show that a
telomeric
DNA sequence, when treated as a quantum wire in the fully coherent
low-temperature regime, works as an excellent semiconductor. 
Clear steps are apparent in the current-voltage curves of telomeric sequences
and are present independent of lengths and sequence initialisation at the
contacts. 
%The current-voltage curves suggest the existence of stepped structures
independent of length and sequencing initialisation at the contacts. 
We also find that the molecule-electrode coupling can drastically influence the
magnitude of the current.
The difference between telomeric DNA and other DNA, such as $\lambda$-DNA and
DNA for the tumour suppressor p53, is particularly visible in the length
dependence of the current. 
\end{abstract}

\maketitle
%%%%%%%%%%%%%%%%%%%%%%%%%%%%%%%%%%%%%%%%%%%%%%%%%%%%%%%%%%%%
%% comment out for single column
%%%%%%%%%%%%%%%%%%%%%%%%%%%%%%%%%%%%%%%%%%%%%%%%%%%%%%%%%%%%
%\narrowtext
%\tighten

%%%%%%%%%%%%%%%%%%%%%%%%%%%%%%%%%%%%%%%%%%%%%%%%%%%%%%%%%%%%
\section{Introduction}
\label{sec:intro}
%%%%%%%%%%%%%%%%%%%%%%%%%%%%%%%%%%%%%%%%%%%%%%%%%%%%%%%%%%%%

Following the publication of the seminal work of Fink and Sch\"{o}nenberger on
the electrical conduction of DNA strands \cite{FinS99} and, shortly afterwards,
a single molecule version by Porath, Bezryadin, Vries and Dekker
\cite{PorBVD00}, the possibilities of electronic transport in DNA have
stimulated a large body of work for more than a decade.\cite{GenB10} This is of
course driven by the exciting fundamental characteristics of DNA as a biological
structure with self-recognition and self-assembly properties.
However, even after all this work, we are still far removed from a proper
understanding of charge transport (CT) in DNA. Experimental work continues to
remain very challenging and publications --- while reporting tantalising hints
of surprisingly large currents --- have only recently  begun to probe CT
properties in DNA beyond the magnitude of the supported
current.\cite{FelSGN08,GuoGH08,SliMRB11} Similarly, the multitude of theoretical
studies has yet to agree on which conduction mechanism and model to choose for a
proper coarse-grained description of CT in DNA --- or to find the means of
attacking the complexities of larger DNA strands via powerful \textit{ab initio}
methods.

Some of the problem stems from the fact that DNA is not just a
single molecule, but rather describes the whole set of DNA strands made possible
by stringing the nucleotide bases Adenine (A), Cytosine (C), Guanine (G) and
Thymine (T) into the classical double helix structure. Consequently, it is
difficult to compare the results of publications when the DNA sequences used are
different, or, as sometimes happens, not fully specified. Furthermore, the used
DNA sequences range from simple periodic poly(dG)-poly(dC) variants --- and of
course their AT counterparts --- to viral, bacterial and eukaryotic DNA and
beyond into artificially generated sequences.

Here, we intend to contribute to the discussion by putting forward a naturally
occurring DNA sequence as a particularly well suited test bed for DNA
experiments as well as theoretical studies: telomeric DNA. In mammals, it is a
Guanine rich sequence in which the pattern TTAGGG is repeated over thousands of
bases. Its length is known to vary widely between species and individuals and it
essentially has a buffer function at the beginning and end of DNA strands for
eukaryote cells.\cite{AlbBLR94} It therefore combines the advantages of a
simple, periodic structure with the richness and biological function of a real
DNA sequence.
In addition, we shall study not a single, but rather $5$ different
coarse-grained, tight-binding models of DNA. In this way, and in the absence of
a
preferred model as discussed above, we shall not be over concerned with
quantitative differences between the models, but rather aim to elucidate their
qualitative agreements.

\begin{figure}[t!]
%\vskip2.0cm
\includegraphics[width=0.95\columnwidth]{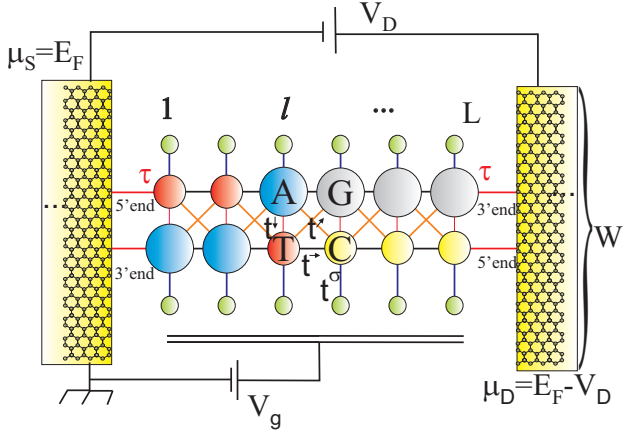}
%(b)\includegraphics[width=0.45\columnwidth]{fig_1.eps}
%\vspace{-0.2cm}
\caption{(color online)  Schematic representation of a ladder-type model for
electronic transport along DNA between two semi-infinite graphene nanoribbon source (S) and
drain (D) contacts as indicated. A third (gate) terminal modulates the conductance
of the molecule. The nucleotide base pairs sequence is indicated by small
(pyrimidines, red, yellow) and large (purines, blue and grey) circles
representing the four possible effective nucleotides, T, C, A, and G,
respectively. The sugar-phosphate backbone is given by green circles and
possible electronic pathways are shown as lines.}
\label{fig_1}
\label{fig:device}
\end{figure}

Some of the most exciting solutions to the well-established contact problem in
nano-devices is based on the use of carbon allotropes such as carbon nanotubes
(CNT).\cite{FelSGN08,GuoGH08} These CNT then allow for good coupling to
macroscopic gold contacts. Even simpler should be the use of graphene flakes or nano-ribbons as
potential contact material. In fact, the basic building block of a DNA device
might very well be a single graphene sheet into which gaps are fabricated by
lithographic techniques. It is these devices which we will take as our starting
point here (cp.\ Fig.\ \ref{fig:device}). 
%%%%%%%%%%%%%%%%%%%%%%%%%%%%%%%%%%%%%%%%%%%%%%%%%%%%%%%%%%%%
%%%%%%%%%%%%%%%%%%%%%%%%%%%%%%%%%%%%%%%%%%%%%%%%%%%%%%%%%%%%

%%%%%%%%%%%%%%%%%%%%%%%%%%%%%%%%%%%%%%%%%%%%%%%%%%%%%%%%%%%%
\section{Theoretical formulation}
\label{sec:theory}
%%%%%%%%%%%%%%%%%%%%%%%%%%%%%%%%%%%%%%%%%%%%%%%%%%%%%%%%%%%%

A schematic figure of a molecule of DNA coupled to graphene electrodes is shown
in Fig.\ \ref{fig_1}. The source and drain electrodes are identified with their
respective Fermi levels $\mu_S$ and $\mu_D$, the effect is taken into account
using self-energy functions.\cite{Datta99} A gate
terminal modulates the conductance of the molecule. We focus on coherent
transport and hence compute $G^r$, the retarded Green's function between the
electrodes S and D via the recursive lattice Green's function
technique.\cite{FG97} This then allows us to calculate the density of state,
transmission
spectrum and the current voltage characteristics.

%%%%%%%%%%%%%%%%%%%%%%%%%%%%%%%%%%%%%%%%%%%%%%%%%%%%%%%%%%%%
\subsection{The model Hamiltonians}

The study of nanoelectronic applications based on macromolecules as building
blocks is a compromise between the model complexity and the length scale that
the
chosen model can handle. Therefore, the use of heuristic tight-binding  models
has grown in interest in the last decade.\cite{CunMRR07} In spite of the simplified
electronic
structure, very long systems can be studied with these models in
contradistinction to \textit{ab initio}
approaches. Hence, we study here the transport properties of DNA
according to 5 such heuristic models. For sake of clarity, the $5$ models
are briefly described as follows. First we address a single strand chain with
$4$ different sites, representing the nucleotides (1L).\cite{liY01} This simplest
possible model
is followed by a single strand of bases attached to sites emulating the
sugar-phosphate backbones (1L+BB or "fishbone" model).\cite{JoeLH10} Next in
complexity appears a double strand chain of nucleotides (2L),\cite{CarLM11} and a
model where the said double strand of base pairs is attached to backbones
(2L+BB)\cite{KloRT05} and, finally, a double strand of base pairs with backbones and
additional diagonal inter chain hoppings among base pairs (2L+BB+D).\cite{WelSR09} 

Instead of repeating the mathematical definitions of the corresponding
Hamiltonians as given in the above cited literature, let us concentrate on the
model 2L+BB+D as the most complicated one. The other $4$ models can then be
reconstructed from it by suitable choice of parameters.
In Fig.\ \ref{fig:device}, we show a schematic representation of 2L+BB+D.
%which includes the full double-stranded structure of the DNA molecule, includes
The model contains parameters for the description of the backbone as well as
longitudinal, transverse and diagonal base to base hopping terms which take
into account the size difference
between pyrimidines (C,T) and purines (A,G). The Hamiltonian can be written as
\begin{equation}
%\label{eq:2channel}
\label{eq:2LBBD}
\begin{array}{l}
H_\text{DNA}=
 \displaystyle\sum_{l=1}^{L}
 %\left[
 \Big[
  \sum_{\alpha=1}^{2}(\varepsilon_{l\alpha}| l,\alpha\rangle\langle l, \alpha |
- t^{\rightarrow}_{l,\alpha}| l,\alpha\rangle\langle l+1,\alpha|)\\ %\right]\\
  -t^{1\searrow 2}_{l}|l,1\rangle\langle l+1,2| - t^{2\nearrow1}_{l}|
l,2\rangle\langle l+1,1| - t^{\downarrow}_{l}| l,1\rangle\langle l,2|\\
  +\displaystyle\sum_{\sigma=\uparrow,\downarrow}(\varepsilon_{l}^{[\sigma]}|
l,\sigma\rangle\langle l,\sigma|-t_{l}^{[\sigma]}| l,\alpha(\sigma)\rangle\langle
l,\sigma |)\Big]
 %\big]
  +\text{h.c.},
\end{array}
\end{equation}
where $t^{\rightarrow}_{l,j}$ is the hopping at base pair $l$ along the strand
starting from 5' ($j=1$) and 3' ($j=2$) ends, $t^{1\searrow 2}_{l}$ and
$t^{2\nearrow1}_{l}$ denote the diagonal hopping and $t^{\downarrow}_{l}$ the
hopping perpendicular from 5' down to 3' at $l$. The sum over $\sigma$ in
\eqref{eq:2LBBD} marks the connection to the sugar-phosphate backbone as
in Fig.\ \ref{fig:device}. Last, the hermitian conjugate indicates the hopping
terms associated with $t^{\leftarrow}_{l,j}=t^{\rightarrow}_{l-1,j}$,
$t^{1\swarrow 2}_{l}=t^{2\nearrow 1}_{l-1}$, $t^{2\nwarrow 1}_{l}=t^{1\searrow
2}_{l-1}$, $t^{\uparrow}_{l}=t^{\downarrow}_{l}$ and $t_{l}^{[\sigma]}$. In
addition, $\alpha(\sigma)=1$ ($2$) for $\sigma=\uparrow$ ($\downarrow$) and
$\epsilon_{l\alpha}$ and $\epsilon^{[\sigma]}_{l}$ denote the onsite energies on
the $2$ DNA strands and the top and bottom backbones, respectively

The different terms are better appreciated by referring to Fig.\ \ref{fig_1}.
The
onsite energies $\epsilon_{l\alpha}$ are taken to be the effective primary
ionization energies  of the base nucleotides, i.e. $\varepsilon_A=8.24$ eV,
$\varepsilon_T=9.14$ eV, $\varepsilon_C=8.87$ eV and $\varepsilon_G=7.75$ eV. In
this work, we consider the  backbone energy to be given as average of the
energies of the base nucleotides, i.e.\ 
$\varepsilon_{l}^{[\uparrow]([\downarrow])}=8.5$ eV for all $l$.
Both strands of DNA and the backbone are modelled explicitly and the different
diagonal overlaps of the larger purines and the smaller pyrimidines are taken
into account by suitable inter-strand
couplings.\cite{WelSR09,RakVMR02} The intra-strand couplings are
$t^{\rightarrow(\leftarrow)}_{l,\alpha}=0.35$ eV between identical bases and
$0.17$ eV
between different bases; the diagonal inter-strand couplings are $t^{1\swarrow
2}_{l}=t^{2\nearrow 1}_{l-1}=t^{2\nwarrow 1}_{l}=t^{1\searrow 2}_{l-1}=0.1$ eV
for
purine-to-purine, $0.01$ eV for purine-pyrimidine and $0.001$ eV for
pyrimidine-to-pyrimidine. Perpendicular couplings to the backbone sites are
$t^{[\sigma]}_{l}=0.7$ eV, and the perpendicular hopping across the hydrogen bond
in a base pair is reduced to $t_{l}^{\downarrow}=0.005$ eV. From previous
discussions leading to these choices of parameters as well as the influence of
the environment on the charge migration properties of the models, we refer the
reader to the existing literature.\cite{Cha07,BerC07,KloRT05,ShiWHC12}
Although we use rather simple Hamiltonians to describe the molecule, we believe
that the qualitative physics of CT in the molecule is captured. This is because
both the molecular energy levels and the wave functions closely resemble those
calculated from the much more sophisticated \textit{ab initio}
theory.\cite{Cha07}
Nevertheless, we emphasise that the choice of the tight binding parameters is
far from uniquely determined, being a rather controversial issue, and several
parameter sets have been proposed in the literature.\cite{Roc03}

%%%%%%%%%%%%%%%%%%%%%%%%%%%%%%%%%%%%%%%%%%%%%%%%%%%%%%%%%%%%
\subsection{Green's function techniques}

The transmission probability $T(E)$ between the electrodes can be evaluated by
\begin{equation}
\label{eq:transmission}
\begin{array}{l}
        T(E)=\text{Tr}\left[ \Gamma_\text{S} G^r\Gamma_\text{D} (G^r)^\dag
\right],
\end{array}
\end{equation}
where $G^r$ is the retarded Green's function of the system which can be found
from\cite{Datta99}
\begin{equation}
\label{eq:GR}
\begin{array}{l}
   G^r=[E \openone-H_\text{DNA}-U-\Sigma_\text{S}-\Sigma_\text{D}]^{-1}.
\end{array}
\end{equation}
Here $\openone$ is the identity matrix, $U$ is the gate potential and
$\Sigma_\text{S}$, $\Sigma_\text{D}$ are the self-energies for source (left) and
drain (right) contacts (cp.\ Fig.\ \ref{fig:device}), respectively.
These self-energies are calculated as usual from $g$, the Green's function of
the electrode,\cite{Datta99} and the coupling $\tau$ between the DNA molecule and
electrode  (cp.\ Fig.\ \ref{fig:device}), i.e.\ 
%\begin{equation}
%label{eq:tau}
%\begin{array}{l}
  $\sum_{\text{S},\text{D}}=\tau_{\text{S},\text{D}}^\dag
g_{\text{S},\text{D}}\tau_{\text{S},\text{D}}$.
%\end{array}
%\end{equation}
The Green's function $g$ is calculated using a recursive
technique.\cite{LSLSR85} The electrode-molecule coupling $\tau$ is determined by
the geometry of the chemical bond.\cite{LarBD97} We use a constant coupling
$\tau=0.35$ eV of similar magnitude as the inter-chain DNA
couplings.\cite{PorBVD00} We emphasise that our results remain robust for small
changes in this parameter.
The anti-Hermitian part of the self-energy is known as the broadening function
and related to the lifetimes $\Gamma_{\text{S},\text{D}}= i
\left(\Sigma_{\text{S},\text{D}}-\Sigma_{\text{S},\text{D}}^\dag\right)$ of an
electron in a
molecular eigenstate.
The H\"{u}ckel approach\cite{Bur84} predicts that the Fermi energy $E_F$ is
closer
to the highest occupied molecular orbital (HOMO) level. In this work the Fermi energy of the Dirac cones in undoped graphene is aligned to the mid gap ($\sim 8.1$ eV) of the DNA model shown in Fig.\ \ref{fig:device}.
%\cite{CasGPN09} 
Consequently, the
Fermi levels in source and drain are $\mu_S=E_F$ and $\mu_D=E_F+e V_{D}$,
respectively.
The gate potential $U$, which allows to model the charging effects in the
presence of bias,\cite{DamRP02} can be expressed as
$U= e V_g$ assuming that there is
no large charge redistribution by applying bias between the electrodes.
%The parameter $(0<\beta<1)$ is a measure of the gate control. In the ideal situation, which we assume here, we have $\beta=1$ and $U$ is essentially tied to the gate voltage. 

Given $H$, $\sum_{\text{S},\text{D}}$, the chemical potentials $\mu_{S,D}$ in
the electrodes and the gate potential $U$, we obtain the density of states DOS
from $G^r$
as
 \begin{equation}
\label{eq:dos}
\begin{array}{l}
   \rho(E)=-\frac{1}{\pi}\text{Tr}\{\text{Im}[G^r(r,E)]\}.
\end{array}
\end{equation}
and the current is given as usual via
$I=- \frac{2e}{h}\displaystyle\int^{\infty}_{-\infty}T(E)\left[
f_S(E)-f_D(E)\right]dE$. 
In the low-temperature limit such that $|e|V_D=\mu_S-\mu_D \gg k_\text{B}T$, we
have $f_{S,D}(E)=
\Theta(\mu_{S,D}-E)$, where $\Theta$ is the step function. So, the electronic
low-$T$ current can be expressed as\cite{Datta99}
\begin{equation}
\label{eq:current}
\begin{array}{l}
  I=-\frac{2e}{h}\displaystyle\int^{\mu_D}_{\mu_S}T(E)dE.
\end{array}
\end{equation}

%%%%%%%%%%%%%%%%%%%%%%%%%%%%%%%%%%%%%%%%%%%%%%%%%%%%%%%%%%%%
\section{Current-voltage characteristics along DNA systems}
\label{sec:iv}
%telemers, p53, l-DNA
%%%%%%%%%%%%%%%%%%%%%%%%%%%%%%%%%%%%%%%%%%%%%%%%%%%%%%%%%%%%

%%%%%%%%%%%%%%%%%%%%%%%%%%%%%%%%%%%%%%%%%%%%%%%%%%%%%%%%%%%%
\subsection{Selection of the DNA sequences}
\label{sec:DNA}

\begin{figure}[t!]
%\vskip2.0cm
%\includegraphics[width=0.95\columnwidth]{Fig_2_telomeric.eps}\\
%\includegraphics[width=0.95\columnwidth]{Fig_2_lambda.eps}\\
%\includegraphics[width=0.95\columnwidth]{Fig_2_p53.eps}
\includegraphics[width=0.95\columnwidth]{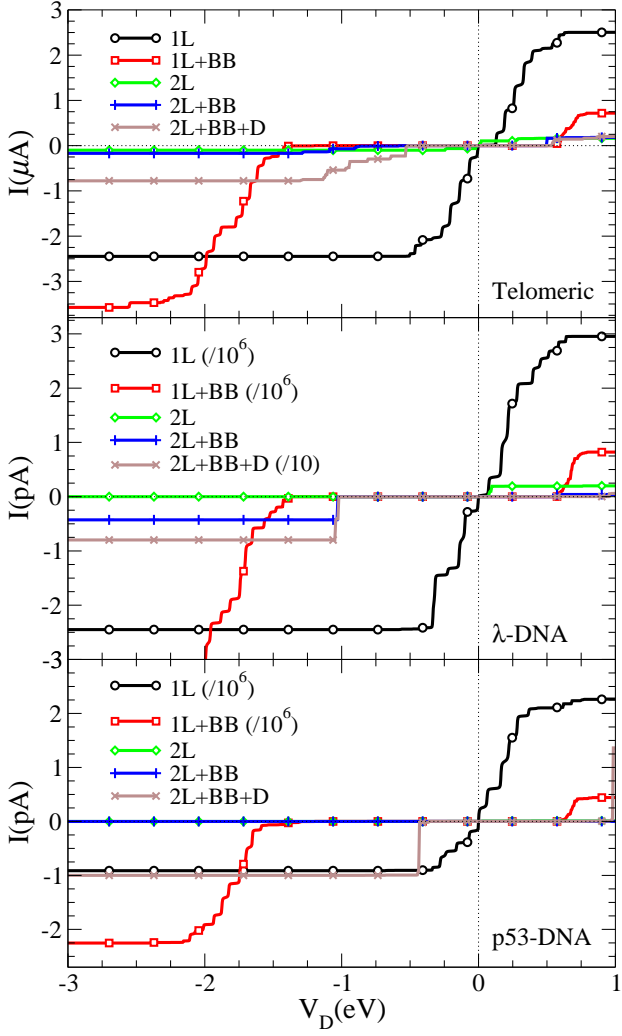}
%\vspace{-0.2cm}
\caption{(color online) I-V characteristic for (top) telomeric, (centre)
$\lambda$-DNA, and (bottom) p53-DNA in $5$ different heuristic models with $L=30$ bps: (i) 1D,
(ii) 1D with backbone (BB), (iii) and (iv) two channels without and with
backbone and (v) two channels with backbone and diagonal coupling. Only every $100$th symbol is shown for clarity. In the
unidimensional case the hopping integrals $t_{i,i+1}$ are assumed to be
nucleotide-independent with $t_{i,i+1}=0.4$ eV. The dotted horizontal and
vertical lines indicate the $I=0$ and $V=0$ constants.}
\label{fig_2}
\label{fig:tel-lam-p53}
\end{figure}

The results shown in Fig.\ \ref{fig:tel-lam-p53} are the calculated I-V
characteristics of three different types of DNA sequences for short DNA strands
of length $L=30$ bps. 
%%%%%%%%%%%%%%%%%%%%%%%%%%%%%%%%%%%%%%%%%%%%%%%%%%%%%%%%%%%%
%%%%%%%%%%%%%%%%%%%%%%%%%%%%%%%%%%%%%%%%%%%%%%%%%%%%%%%%%%%%
The sequences considered are (i) the eukaryotic telomer based on repeats of the
TTAGGG sequence as discussed in in section \ref{sec:intro},\cite{AlbBLR94,Zak95} (ii) a
random subsequence of bacteriophage $\lambda$-DNA\cite{FogAS10,dePMHCGHHBOSA00} and (iii) a random
subsequence of the DNA strand of the p53 gene.\cite{TraAG00,ShiRR08} The
$\lambda$-DNA sequence has been studied previously as a typical example of a
biological DNA sequence. It contains $48502$ base pairs with $25\%$ of A, $24\%$
of C, $28\%$ of G, and $23\%$ of T strung together in a non-periodic sequence.
Similarly, p53-DNA has $20303$ base pairs and is part of an important regulatory
mechanism in humans.\cite{She04,TubT11,MalHS10}

%%%%%%%%%%%%%%%%%%%%%%%%%%%%%%%%%%%%%%%%%%%%%%%%%%%%%%%%%%%%
\subsection{Model dependence of the current-voltage characteristics}
\label{sec:model}

\begin{figure}[t!]
%\vskip2.0cm
\includegraphics[width=0.95\columnwidth]{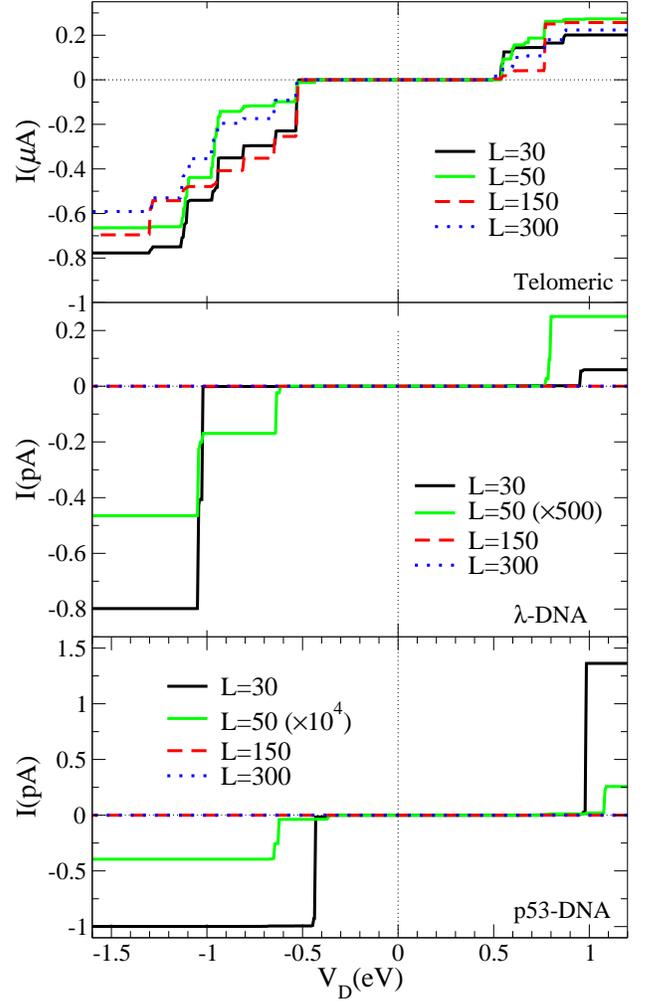}
%\includegraphics[width=0.45\columnwidth]{Fig_3b.eps}
%\vspace{-0.2cm}
\caption{(color online) I-V characteristic of DNA segments with $L=30$, $50$, $150$  and
$300$ bps for (top) telomeric, (centre) $\lambda$ and (bottom) p53-DNA sequences.
The thin dotted horizontal and vertical lines indicate the $I=0$ and $V=0$
constants.}
\label{fig_3}
\label{fig:lengthdependence}
\end{figure}

The main results for the I-V characteristics depicted in Fig.\
\ref{fig:tel-lam-p53} can be summarized in a few general trends. 
First, the inclusion of backbones opens a gap between the HOMO and the
lowest unoccupied molecular orbital (LUMO), clearly revealed in the
corresponding I-V curves showing
threshold voltages.\cite{CCPD02} Here it is worth recalling that we assume a
Fermi energy close to half the energy gap, by tuning a gate voltage of $V_g=9.8$ eV for all models (see appendix \ref{sec:gatevoltage}). For sufficiently high
applied voltages the entire band will be scanned across the Fermi energy,
leading to current saturation, as can be observed for almost all models for the
voltage range depicted in Fig.\ \ref{fig:tel-lam-p53} and further illustrated in
the appendix. 
Next, a non-periodic sequence of base pairs, such as in $\lambda$- and p53-DNA,
can drastically suppress the currents, even for such a short system of $L=30$
 bps, as one sees comparing Fig.\ \ref{fig:tel-lam-p53}, for disordered
$\lambda$-DNA and p53-DNA, respectively, to the telomer sequence. This strong suppression is model dependent: single strands do not show a
large supression at such short lengths, while double strand models already
exhibit a six orders of magnitude drop in current. 
Furthermore, the
double stranded chain 2L+BB+D, i.e.\ with diagonal hopping and backbones, shows
higher currents
than the other double stranded models, partially recovering single strand
values. 
Last, and important in what follows, there are steps appearing in some I-V
curves, which
are due to resonances in the transmission probabilities. Such resonances are
more relevant and robust in the telomeric sequences, due to the split of each
band in several sub-bands.\cite{KloRT05} Resonance effects are still present in
some cases for
disordered short sequences, but they do not last for longer systems, as can be
systematically appreciated in Fig.\ \ref{fig:lengthdependence}. 
%%%%%%%%%%%%%%%%%%%%%%%%%%%%%%%%%%%%%%%%%%%%%%%%%%%%%%%%%%%
%%%%%%%%%%%%%%%%%%%%%%%%%%%%%%%%%%%%%%%%%%%%%%%%%%%%%%%%%%%
Hence we see that the DNA model proposed in \eqref{eq:2LBBD} seems to
interpolate between the extreme cases of the simple 1L models, retains the gap
due to the backbone, but has the smaller current values, that are more in line
with the experimental results. In what follows, all the results are for the
model \eqref{eq:2LBBD}, namely 2L+BB+D, although it is the most complicated one
of those considered previously.

%%%%%%%%%%%%%%%%%%%%%%%%%%%%%%%%%%%%%%%%%%%%%%%%%%%%%%%%%%%%
\subsection{Length dependence of the current-voltage characteristics}
\label{sec:length}

One of the most striking differences between the telomeric and the
non-telomeric
sequences is revealed by comparing short and longer sequences.
Fig.\ \ref{fig:lengthdependence} shows the I-V characteristic for all three DNA
strands, considering in each case four different strand lengths. 
The degree of non-periodicity in the sequences severely affects
the I-V characteristics. First, there is the expected drop in the
current due to non-periodicity (acting like an effective disorder), and, the robustness of the step-like structures in the I-V curves for the telomeric strands at any $L$. Most prominent, however, is the rapid drop in the current when increasing $L$ for the non-telomeric sequences. 
For $300$ bps ($104$ nm), $\lambda$-DNA and p53-DNA, which are 
non-periodic,
support practically no current, even in the $p$A scale, while the telomeric
sequence retains a maximum saturation current of around $0.6$ $\mu$A. 
 
\begin{figure}[t!]
%\vskip2.0cm
%\includegraphics[width=0.45\columnwidth]{Fig_3a.eps}
\includegraphics[width=0.95\columnwidth]{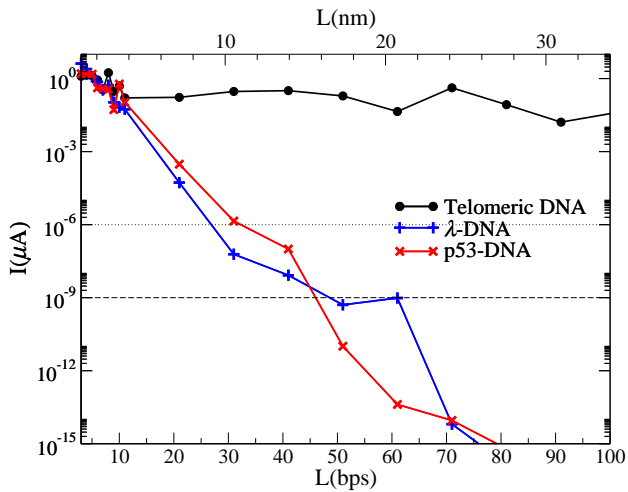}
%\vspace{-0.2cm}
\caption{(color online) Current $I$ as a function of length $L$ for telomeric
($\bullet$), $\lambda$-DNA ($+$) and p53-DNA ($\times$) sequences at $V_D=1.5$ eV. The thin horizontal dotted line corresponds to a current level of $10^{-12}$ A, which can be measured relatively easily, whereas the dashed line at $10^{-15}$ A indicates limits of standard current measurement techniques.}
\label{fig_4}
\label{fig:lengthdependence1_5volt}
\end{figure}

Fig.\
\ref{fig:lengthdependence1_5volt} shows the current of the three
sequences as a function of length at fixed saturation current $V_D=1.5$ eV. For very
short lengths up to
$L \approx 10$ bps, the three cases can not be distinguished. However, for
longer strands, the telomeric sequence shows a saturation
current of the order of several hundreds of $n$A almost constant in $L$, whereas
for $\lambda$-DNA and p53-DNA the current decreases exponentially as the
electronic states are strongly localized.
%%%%%%%%%%%%%%%%%%%%%%%%%%%%%%%%%%%%%%%%%%%%%%%%%%%%%%%%%%%
%%%%%%%%%%%%%%%%%%%%%%%%%%%%%%%%%%%%%%%%%%%%%%%%%%%%%%%%%%%
It should be noted here that this result is robust and independent from the
heuristic 2L+BB+D model, i.e.\ it happens in all the models considered. It
should therefore be advantageous to consider the naturally occurring telomeres of DNA
sequences as prime candidates when looking for good conductivity in a
DNA strand --- in nanotechnology applications but also in DNA-related in-vivo processes where charge migration might be important.\cite{BerC07}

%%%%%%%%%%%%%%%%%%%%%%%%%%%%%%%%%%%%%%%%%%%%%%%%%%%%%%%%%%%%
\section{Transport characteristics of telomeric DNA}
\label{sec:telomers}
%%%%%%%%%%%%%%%%%%%%%%%%%%%%%%%%%%%%%%%%%%%%%%%%%%%%%%%%%%%%

In order to establish the features of the electronic transport which are
intrinsic to the telomeric sequence, the role of the possible variations in
coupling to the graphene contacts has to be studied. Three main aspects are
considered: (i) the dependence on which base pair of the telomeric sequence is
connected to the graphene, (ii) whether contacts are only made to a single
strand of the DNA or to both, 3' and 5', ends, and (iii) what happens when more
than a single telomer is contacted in parallel. 
Let us recall that for the results presented up to now, we always assumed that
the transport starts at the T of the TTAGGG sequence, that the coupling $\tau$
holds for both 3' and 5' ends and that only a single telomer has been contacted.

%%%%%%%%%%%%%%%%%%%%%%%%%%%%%%%%%%%%%%%%%%%%%%%%%%%%%%%%%%%%
\subsection{Starting the telomer}
\label{sec:startpoint}

\begin{figure}[t!]
%\vskip2.0cm
%\includegraphics[width=0.65\columnwidth]{Fig_4a.eps}
%\includegraphics[width=0.45\columnwidth]{Fig_4_TTAGGG_inv.eps}
%\includegraphics[width=0.65\columnwidth]{Fig_4_TTAGGG.eps}
%\includegraphics[width=0.45\columnwidth]{Fig_4_TAGGGT.eps}
%\includegraphics[width=0.65\columnwidth]{Fig_4_AGGGTT.eps}
%\includegraphics[width=0.45\columnwidth]{Fig_4_TAGGGT-AGGGTT.eps}
%\includegraphics[width=0.95\columnwidth]{Fig_4_IvsV.eps}
\includegraphics[width=0.95\columnwidth]{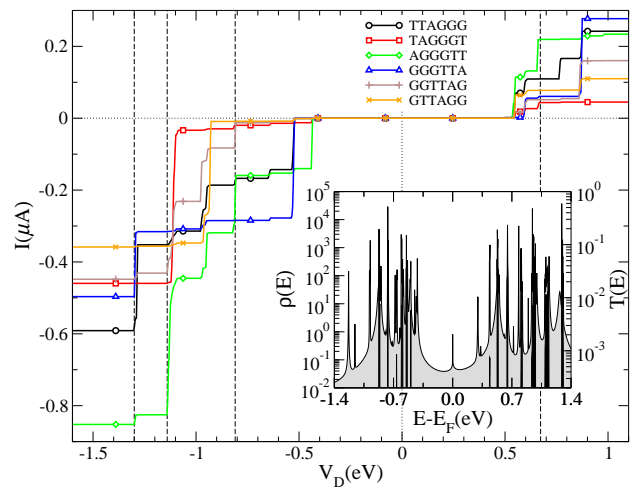}

%\vspace{-0.2cm}
\caption{(color online) 
I-V characteristic for telomeric sequences ($L=300$ bps long) and differently
attached to the contacts. The inset shows how the sequence starts at the left
contact in each case, with the first letter connected to the graphene edge. The dashed vertical lines mark, for example, four positions of very robust steps (see text). $I=0$ and $V=0$ constants are also indicated as in Fig.\ \ref{fig:lengthdependence1_5volt}. Only every $100$th data point is shown for clarity.
Inset: DOS (grey shaded black line) and $T(E)$ (black line) for the telomer TTAGGG attached to the graphene contacts.}
\label{fig_5}
\label{fig:startpoint}
\end{figure}

In Fig.\ \ref{fig:startpoint}, we show I-V curves for telomeric sequences
starting at different base pair positions, i.e.\ in addition to the periodic
repeat of TTAGGG, we also have TAGGGT, AGGGTT, \ldots, GTTAGG. In this way, we
hope to capture the possibilities which might arise when the contacts are made
simply by putting telomers on top of graphene sheets.\footnote{More
sophisticated, wet-chemical methods can of course be used to assure contacts at
selected bases only.\cite{FelSGN08,RoyVRK08,GuoGH08}} 
The DOS of the TTAGGG system, as well as $T(E)$, is also shown in Fig.\
\ref{fig:startpoint}. The results are for $L=300$ bps in all cases. 
%%%%%%%%%%%%%%%%%%%%%%%%%%%%%%%%%%%%%%%%%%%%%%%%%%%%%%%%%%%%
%%%%%%%%%%%%%%%%%%%%%%%%%%%%%%%%%%%%%%%%%%%%%%%%%%%%%%%%%%%%
The DOS shows a peak at $E_F$ related to the nanoribbon localized state at this
energy. The overall envelope of the DOS is shaped by the graphene band structure
and the band splitting of the
DNA system. The superposed sharp peaks are related to the resonances in
$T(E)$, which show a huge variation in intensity. Therefore, some peaks in
the DOS are related to transmission peaks that fall below the scale of the
figure.
The irregular step-like structure in the I-V
characteristics reflects the fragmented structure of $T(E)$.
Notice that the apparent gap in the I-V characteristics does not necessarily
correspond to the gap in the DOS. The reason is that many of the states close to
the band gap have very low transmission probabilities (related to highly
localized resonances at different parts of the rather long sequences), therefore
they do not contribute to transport. 

When inspecting the I-V characteristics for the different starting positions,
one notices that each one shows steps at the \emph{same} energy, although the
steps show variable amplitudes. We therefore find that, while changing the start
of the sequence at the contact can alter the height of the transmission
probability  for a given energy, it does not change the position of the steps in
the I-V curves. As we had argued in section \ref{sec:iv}, it is exactly the
existence of these pronounced steps which makes the telomeric sequence stand out
from other DNA sequences. Our result of the section now show that this finding
remains robust with respect to what actual base pair is chosen to contact to the
graphene sheets.

%%%%%%%%%%%%%%%%%%%%%%%%%%%%%%%%%%%%%%%%%%%%%%%%%%%%%%%%%%%%
\subsection{Contacting the telomer at 3', 5' or both}
\label{sec:SSvsDS}

\begin{figure}[t!]
%\vskip2.0cm
\includegraphics[width=0.95\columnwidth]{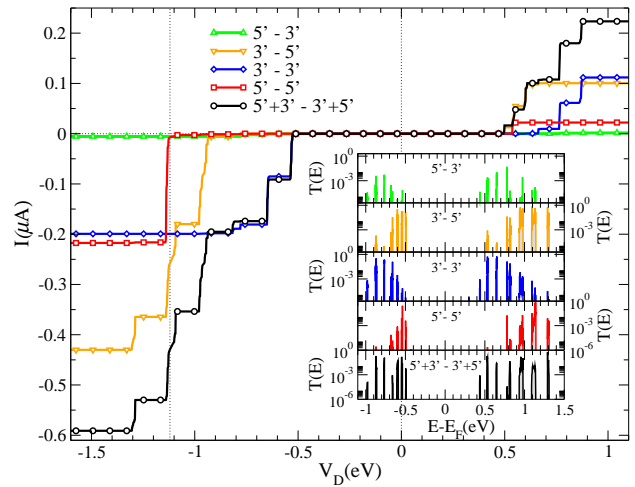}
%\includegraphics[width=0.95\columnwidth]{Fig_5a.eps}
%\includegraphics[width=0.95\columnwidth]{Fig_5b.eps}
%\vspace{-0.2cm}
\caption{(color online) I-V characteristics for the finite DNA ladder, $300$ bps
long, contacted to electrodes in five different ways (cp.\ Fig.\ \ref{fig:device}): source 5' to drain 3` (green $\triangle$), source 3` to drain 5` (orange $\triangledown$), source 3` to drain 3` (blue $\diamond$), source 5` to drain 5` (red $\square$) and last, both source 5' and 3' to the drain 3' and 5' (black $\circ$). The vertical dotted line denotes a robust step in the I-V curves for all contact geometries. $I=0$ and $V=0$ constants are also indicated as in previous figures.
Inset: $T(E)$ for the five different contact situations with the same colours as in the main figure.}
\label{fig_6}
\label{fig:connections}
\end{figure}

In the literature,\cite{FelSGN08,GuoGH08} it is known that the choice of
contacting single 3' and 5' ends only or both 3' and 5' ends can alter the
current response. This is also observed in our 2L+BB+D model --- the class of 1L
models can of course not capture this behaviour --- as we show in Fig.\
\ref{fig:connections}.
%%%%%%%%%%%%%%%%%%%%%%%%%%%%%%%%%%%%%%%%%%%%%%%%%%%%%%%%%%%%
%%%%%%%%%%%%%%%%%%%%%%%%%%%%%%%%%%%%%%%%%%%%%%%%%%%%%%%%%%%%
In the figure, we have contacted the graphene electrodes via (i) only the direct 3'-to-5' or 5'-to-3' ends at source and drain,
(ii) only the cross 3'-to-3' or 5'-to-5' ends and (iii) both 3' and 5' ends. Besides the trivial effect of dropping the current by
reducing the number of channels, as readily observed by comparing the black
curve to the others, one also sees that all cases shows a significant modification of the sub band transmission
intensities in the energy range of interest around the Fermi energy. 
We interpret the difference between 5' and 3' results in Fig.\
\ref{fig:connections} as being due to the sequence of nitrogenous bases.  In the source 5' terminal, the onsite energy for the thymines ($9.14$ eV) is larger than the onsite energy for the adenine ($8.24$ eV) in the 3' terminal. So, a contact via the A in 3' is preferable. In addition, direct transmission as in 3'-to-5' can make use of the larger hopping strengths compared to the diagonal or perpendicular couplings. We note, however, that this does not fully explain the sequence of current strengths, e.g.\ it is not immediately clear why the source 5' to drain 5' current is larger than the source 5' to drain 3' current.
Nevertheless, the present
results strongly suggest that the step structure of the I-V characteristics,
although intrinsic to the telomeric sequence DOS, are filtered and selected by
the choice of how the strands are coupled to the contacts. Note that we
still observe that the (voltage) position of the steps remains fairly robust.

%%%%%%%%%%%%%%%%%%%%%%%%%%%%%%%%%%%%%%%%%%%%%%%%%%%%%%%%%%%%
\subsection{More than a single telomer in parallel}
\label{sec:parallel}

The advantage of the DNA device outlined in section \ref{sec:intro} and shown in
Fig.\ \ref{fig_1} is the macroscopic size of the graphene contacts together with a small,
nano-sized gap to be manufactured into the graphene sheet. It is then 
clear that it will be hard to guarantee that only a single DNA strand lies
across the gap. Hence we expect that one might occasionally encounter a
situation where more than one DNA strand is being contacted. As we shall show
here, this seems indeed beneficial, i.e.\ we find that the situation of
telomeric DNA strands contacted in parallel enhances the size of the current.

\begin{figure}[t!]
%\vskip2.0cm
\includegraphics[width=0.95\columnwidth]{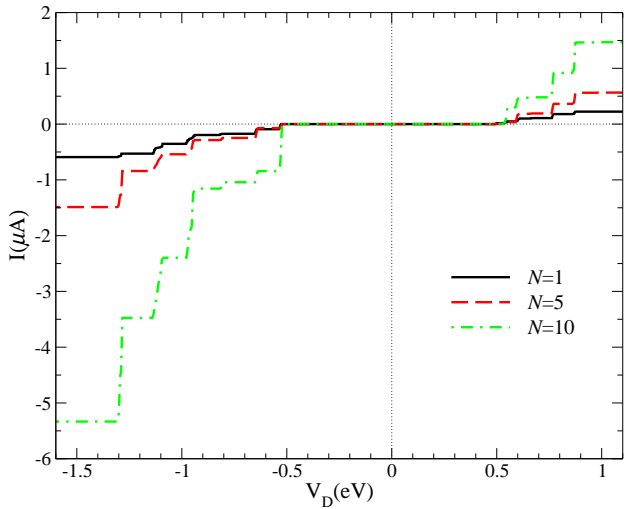}
%\vspace{-0.2cm}
\caption{(color online) I-V characteristic for telomeric DNA arranged as $N=1$,
$5$ and $10$ parallel strands, each of which is $301$ bps
long.}
\label{fig_7}
\label{fig:parallel}
\end{figure}

In Fig.\ \ref{fig:parallel}, we show the I-V characteristics for a single
sequence, together
with the curves for $5$ and $10$ sequences in parallel, all of them $L=301$ bps
long, and separated each by $40$ C-C distances on the graphene contacts.
We start the first sequence with TTAGGG\ldots and hence end after $301$ bps as \ldots TTAGGGT. Then we use the remainder of the telomer as the start sequence of the next strand, i.e.\ TAGGGT\ldots AGGGTT. The end of the $5$th strand is then \ldots GTTAGG and the $10$th strand ends as \ldots GGGTTA. In this way, we have different starting and end parts of the telomer at the source and drain contacts, respectively, for parallel sequences.
From Fig.\ \ref{fig:startpoint}, we can
see that the step structure is indeed enhanced and not washed out by
adding many strands, even if their assembly in an actual experiment is
not base pair exact. It should also be noticed that the currents are not a
simple linear
addition of the $N=1$ case, since interference effects appear, due to the
different sequences coupled to the same contacts. 
%%%%%%%%%%%%%%%%%%%%%%%%%%%%%%%%%%%%%%%%%%%%%%%%%%%%%%%%%%%%
%%%%%%%%%%%%%%%%%%%%%%%%%%%%%%%%%%%%%%%%%%%%%%%%%%%%%%%%%%%%
Also, the position of the steps between plateaus remains again very robust.

%%%%%%%%%%%%%%%%%%%%%%%%%%%%%%%%%%%%%%%%%%%%%%%%%%%%%%%%%%%%
\section{Conclusions}
\label{sec:conclusions}
%%%%%%%%%%%%%%%%%%%%%%%%%%%%%%%%%%%%%%%%%%%%%%%%%%%%%%%%%%%%

The complexity of the DNA systems does not allow yet a definitive
conclusion to be drawn on the mechanism(s) leading to charge transport in telomeric or other
sequences. Nevertheless, the consistency among the results found here for
different heuristic models, widely discussed in the literature, suggests that
important qualitative physical and chemical aspects have been captured in the
present approach. Telomeric sequences contacted to graphene sheets may show high
currents as well as robust plateau structures in the I-V characteristics. This
plateau structure is enhanced and not washed out by adding parallel telomeric
chains, irrespective of the starting points defined at the interface of each
sequence with the contact. Most importantly, the drastic difference in the lengths dependence of the current for telomeric DNA as compared to other DNA strands such as the $\lambda$- and p53-DNA shown in Fig.\ \ref{fig:lengthdependence1_5volt} will allow for better comparison and interpretation of experimental data. We emphasise that the inclusion of some external disorder, be it in the fidelity of the sequence or the environmental conditions, does not drastically alter these differences.\cite{KloRT05}
An important current rectification behaviour can also
be seen due to the differences in the transmission probabilities of occupied and
unoccupied levels. A device including a gate electrode could further tune the
rectification possibilities addressed here. 

The devices investigated in the present work --- based on the naturally
occurring, physiochemically stable telomers --- should provide ideal benchmark
situations for
systematic investigations of DNA electronic transport, as well as for
development of DNA based molecular nanoelectronic applications using graphene as
contacts. In particular, the I-V characteristics presented here show promise for a
wide range of interesting nanocircuitry on complex patterns of hollowed graphene
sheets bridged by telomeric DNA sequences.

%%%%%%%%%%%%%%%%%%%%%%%%%%%%%%%%%%%%%%%%%%%%%%%%%%%%%%
%\acknowledgments
%%%%%%%%%%%%%%%%%%%%%%%%%%%%%%%%%%%%%%%%%%%%%%%%%%%%%%
\begin{acknowledgments}
%We thank Neil Wilson for his many instructive and encouraging comments on this
%work and in particular to his suggestions of using graphene contacts as possible
%next generation devices.
%
C.J.P.\ acknowledges financial support from FAPESP (Brazil) and the hospitality
of the Centre for Scientific Computing at the University of Warwick (U.K) were
most of the work has been developed. P.A.S.\ acknowledges partial support
from CNPq (Brazil).
\end{acknowledgments}

\appendix

%%%%%%%%%%%%%%%%%%%%%%%%%%%%%%%%%%%%%%%%%%%%%%%%%%%%%%%
\section{Varying the gate voltage}
\label{sec:gatevoltage}
%%%%%%%%%%%%%%%%%%%%%%%%%%%%%%%%%%%%%%%%%%%%%%%%%%%%%%%

%%%%%%%%%%%%%%%%%%%%%%%%%%%%%%%%%%%%%%%%%%%%%%%%%%%%%%%%%%%%
\begin{figure}[tb]
%\vskip2.0cm
%\includegraphics[width=0.45\columnwidth]{Fig_7a.eps}
%\includegraphics[width=0.45\columnwidth]{Fig_7_vg0.eps}
%\includegraphics[width=0.95\columnwidth]{Fig_7_vg4.eps}
%\includegraphics[width=0.95\columnwidth]{Fig_7_vg9-8.eps}
%\includegraphics[width=0.95\columnwidth]{Fig_7_vg8.eps}
%\includegraphics[width=0.95\columnwidth]{Fig_7_vg6.eps}
\includegraphics[width=0.95\columnwidth]{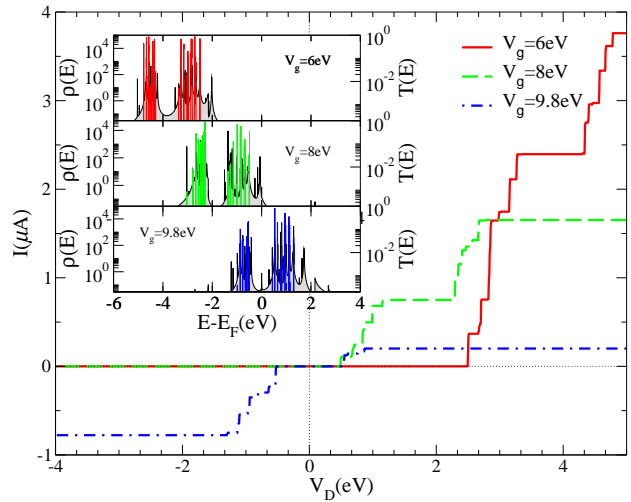}
%\vspace{-0.2cm}
\caption{(color online) I-V characteristics for telomeric DNA using ideal gate
control ($\beta=1$), $L=30$ bps and three different $V_g$:$V_g=6$ eV
(red),
$V_g=8$ eV (green) and $V_g=9.8$ eV (blue). Inset: Density of states $\rho$ (grey shaded) and
transmission probability $T(E)$ (same colours as in main panel)
for the different gate voltages.}
\label{gate}
\end{figure}
%%%%%%%%%%%%%%%%%%%%%%%%%%%%%%%%%%%%%%%%%%%%%%%%%%%%%%%%%%%%
Some features in the I-V characteristics depend on the gate voltage,
$V_g$, applied to the system and are worth briefly discussing here. The
gate voltage will shift the DOS modifying drastically the transport
characteristics. Keeping in mind that the main results shown in the present paper
are for $V_g=9.8$ eV (bottom panel of Fig.\ \ref{gate} inset, showing the
transmission
probability bands superposed to the density of states) a situation where the
equilibrium Fermi
energy lies closer to the HOMO (top of the valence band in semiconductor
terminology), one observes that  negative $V_D$ show higher currents than
positive $V_g$ in Fig.\ \ref{gate}. 
This asymmetry is due to the
lower transmission probability for the occupied molecular orbitals, compared to
the unoccupied molecular orbitals. One can also appreciate how
the saturation currents in the I-V characteristics are related to the energy
ranges beyond the transmission bands. Diminishing the gate voltage, as the
examples shown in the Fig.\ \ref{gate} inset: $V_g=8$ eV (center panel) and
$V_g=6$ eV
(top panel) it is possible to reach a situation of no current for negative
$V_D$, as can be seen for the corresponding I-V characteristics in the Fig.\
\ref{gate}.

%%%%%%%%%%%%%%%%%%%%%%%%%%%%%%%%%%%%%%%%%%%%%%%%%%%%%%%%%%%%
\section{Modelling the graphene contacts}
\label{sec:contacts}
%%%%%%%%%%%%%%%%%%%%%%%%%%%%%%%%%%%%%%%%%%%%%%%%%%%%%%%%%%%%

\begin{figure}[t]
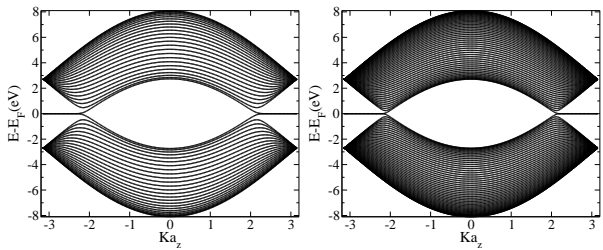

\includegraphics[width=0.45\columnwidth]{Fig_9a.eps}
\includegraphics[width=0.45\columnwidth]{Fig_9b.eps}
\caption{Energy bands for the graphene electrodes with widths $W=25$ (left) and $50$ (right) C-C bond lengths. }
%\label{fig_5}
\label{fig:graphenecontacts}
\end{figure}

The contacts to the DNA devices are semi-infinite graphene nanoribbons as
depicted in Fig.\ \ref{fig_1}. Graphene is a single layer of carbon atoms packed in a
honeycomb lattice, constituted by two sub lattices. Finite pieces of graphene
are limited by two different kinds of edges (or, more generally, a mixing of
both kinds): armchair-like edges, where the atoms at the edges alternately
belong to both sub lattices and a zig-zag edge, in which all the edge atoms
belong to the same sub-lattice. The graphene contacts considered here are
semi-infinite nanoribbons along the zig-zag direction, hence the DNA strands
are connected to an armchair edge (perpendicular to the zig-zag direction) of
width $W$. The thinnest nanoribbons considered are $W=50$ wide, in units of C-C
bond length of $0.142$nm. Bearing in mind that DNA chains exist in many different
conformations that an average measure between $2.2$nm to $2.6$nm wide, the
connection of an increasing number of telomeric DNA double strands lead 
to the use of contacts, up to $W=500$, in units of C-C bond lengths, in the
case of $10$ telomeric DNA strands in parallel.
Fig.\ \ref{fig:graphenecontacts} shows the energy bands of grapheme nanorribons along the zig-zag
direction for two different widths, $W=25$ and $W=50$. 
%%%%%%%%%%%%%%%%%%%%%%%%%%%%%%%%%%%%%%%%%%%%%%%%%%%%%%%%%%%%
%%%%%%%%%%%%%%%%%%%%%%%%%%%%%%%%%%%%%%%%%%%%%%%%%%%%%%%%%%%%
As can be seen here, from
the electronic properties point of view, zig-zag like nanoribbons are metallic
with the valence and conduction band touching at two points as expected (armchair edged nanoribbons
may be semiconductor or metallic, depending on the width).
\cite{CasGPN09,NAkFD96} The flat bands at the Fermi energy are localized
states at the semi-infinite zig-zag edges, giving rise to the peak in the DOS at $E=E_F$ in the inset
of Fig.\ \ref{fig:startpoint}, but showing no influence on the transport properties of the contacted
DNA strands.
Note also that we have obtained qualitatively similar results ---in terms of differences between telomers and $\lambda$- and p53-DNA as well as the lengths dependence of current magnitudes and the robustness of current steps --- when using simple square lattice contacts.

%%%%%%%%%%%%%%%%%%%%%%%%%%%%%%%%%%%%%%%%%%%%%%%%%%%%%%%%%%%%
% RAR automatic references, to be merged later
\bibliographystyle{prsty}
\bibliography{bibliograph}
%%%%%%%%%%%%%%%%%%%%%%%%%%%%%%%%%%%%%%%%%%%%%%%%%%%%%%%%%%%%
\end{document}